# Continuum and Molecular Modeling of Chemical Vapor Deposition over Nano-scale Substrates

**Himel Barua (1) and Alex Povitsky (x)**

(x) Corresponding author: povitsky@uakron.edu, professor

Department of Mechanical Engineering, The University of Akron, Akron OH USA 44325-390

(1) Graduate student, Department of Mechanical Engineering, College of Engineering at the University of Akron, Akron OH 44325-3903

Present affiliation: Post-doctoral research associate at The U.S. Department of Energy's (DOE) Oak Ridge National Laboratory (ORNL); Email: baruah@ornl.gov

5200, 1 Bethel Valley Rd, Oak Ridge, TN 37830

# ABSTRACT

Chemical vapor deposition (CVD) is a common industrial process that incorporates a complex combination of fluid flow, chemical reactions, and surface deposition. Understanding CVD processes requires rigorous and costly experimentation involving multiple spatial scales, from meters to nano-meters. Numerical modeling of deposition over macro-scale substrates has been conducted in literature and results show compliance with experimental data. For smaller scale substrates, where the corresponding Knudsen number is larger than zero, continuum modeling does not provide with accurate results that calls for implementation of molecular-level modeling techniques. In the current study the finite-volume method (FVM) and direct simulation Monte Carlo (DSMC) method have been coupled to model the reactor-scale flow with CVD around micro- and nano- scale fibers.

CVD at fibers with round cross-section is modeled where fibers are oriented perpendicularly with respect to the feedstock gas flow. The DSMC method has been applied to modeling flow around the matrix of nano-scale circular individual fibers. Results show that for smaller diameters of individual fibers with the same filling ratio, the residence time of gas particles inside the fibrous media reduces, and, consequently, the amount of material surface deposition decrease. The sticking coefficient on fibers' surface plays an important role, for instance, increasing the sticking coefficient from 20% to 80% will double the deposition rate.

1. Introduction

Chemical vapor deposition (CVD) is a material deposition process with vapor-to-solid chemical reaction at rigid surface. The CVD is generally used in industrial processes of manufacturing of multi-layer materials to create a thin film over a substrate surface [*1*]. Fluid flow through and around bundles of fibers need to be modeled to obtain the deposition rate at fibers' surface. The set-ups of deposition substrate surfaces vary as the surfaces could be smooth or rough, represent isolated fibers or collective surface of bundles of fibers [*2*]. The scope of the study is to model the CVD of carbon at fibers' surface (Fig. 1) that is qualitatively similar to filtration and physical vapor deposition processes in which solid particles deposit at the fiber surface. Depending on the shape and size of the substrate, the CVD process involves complex no-slip or slip fluid flow phenomena along with material deposition, which makes the fluid-solid system quite complicated to model and to understand.

Though for macro length scales traditional Newtonian and continuum description of fluid mechanics with no-slip boundary conditions can resolve the gas flow phenomena, for a smaller length scale the major discrepancy [*3*] occurs between continuum approximation and real molecular flows that causes substantial variation in CVD rate. To understand the validity of continuum approximation, a well-established non-dimensional group known as the Knudsen number (Kn) is used. The Kn represents the degree of rarefaction of gas by the ratio of the gas molecular mean free path over the geometric length scale of substrate :

$$Kn = \frac{\lambda}{L} \qquad (1)$$

where λ is the gas molecular mean free path and L is the length scale of the system such as diameter of an individual fiber or of bundle of fibers.

In general, gas micro flows encounter compressibility, and viscous heating. In the current case, Ma << 0.3, and therefore, compressibility and aerodynamic (viscous) heating are not substantial.

For a macro-scale continuum flow (Kn<0.01), solving Navier-Stokes equations with no-slip boundary conditions produces solutions coinciding fairly with experiments. For smaller linear scales typical for cross-flow oriented long fibers of micron-scale diameters (0.01<Kn<0.1), the above approach becomes less accurate and for larger Knudsen numbers (Kn>0.1) the no slip approach is an inadequate.

In vapor deposition process, gas rarefaction effect creates a thin flow layer near the rigid surface named the Knudsen layer, which is of the linear scale of a single molecular mean free path. The Knudsen layer cannot be analyzed by using Navier-Stokes equations and, generally speaking, requires solution of Boltzmann equation for local velocity distribution of molecules. For Kn<0.1, the Knudsen layer covers less than 10% of the channel height for internal flows. For Kn<0.1, the Knudsen layer can be neglected by extrapolating continuous bulk gas flow up to the rigid surface, which results in a finite slip velocity at the wall. This flow regime is called partial slip flow regime.

For larger Kn numbers, above continuum approach is no longer accurate and solution of flowfield in Knudsen layer is required that can be achieved by molecular Direct Simulation Monte Carlo (DSMC). In the current research, a methodology has been developed to approach problems with large range of linear flow scales, from reactor scale to individual fibers of sub-micron diameter. Two well-established methods, namely, continuum Finite volume method (FVM) for larger linear scales and molecular Direct Simulation Monte Carlo (DSMC) for flow near fibers have been combined and one-way coupled.

To approach the problem of determining the slip velocity in the current study, the fluid motion has been considered at a molecular scale at least for a part of domain instead of considering the fluid as a continuous medium. At molecular scale, the methods that have been used by researchers are categorized as Molecular dynamics (MD) and Direct simulation Monte Carlo (DSMC). When using MD, every computational particle represents an individual fluid molecule. This approach is more applicable for liquids at nano-scale where molecules interact with their neighbors at all times through potential inter-molecular forces. The DSMC allows a fraction of needed computing time compared to MD as DSMC particles represents billions of physical gas molecules which interact with each other only during their collision. Statistical sampling is used to obtain fluid dynamics properties from DSMC results.

Considerable amount of research has been done to model CVD process. While CVD modeling over flat plate is considered in numerous papers, for example, [4], [5], CVD over more complex and multi-scale geometry drew much less attention [6]. Most of the published approaches model kinetics of chemical reactions while the link between CVD and fluid dynamics remains unexplained. Modeling of coating over a single fiber surface has been investigated [7] in which an adaptive finite element method is used to capture the thickness of deposited material. Ref. [8] conducted experiment and continuous CFD modeling to investigate the growth process of multi-walled carbon nanotube by CVD. In Ref [8], the authors used FVM implemented in FLUENT and studied the flow properties. Most of modeling of Carbon deposition on fibers [9], [10], [11] is limited to either flow modeling around a single fiber [12] or a reactor bulk flow modeling [13], [14]. In the current research, effects of multiple fibers, which are forming arrays of fibers, have been investigated to discover their critical role in spatial and temporal deposition profile and the final thickness of deposited layer of material.

DSMC has been implemented in prior studies [15], [16]. Ref [17] has combined FVM with DSMC to model chemically neutral flow around fibrous medium. Refs. [18], [19] has extended DSMC to model adsorption and desorption in porous media. Ref. [20]adopted total collision energy and Quantum kinetic model [21] to implement chemical reactions in DSMC method. Current research implemented an extension of DSMC method to solve Carbon deposition over wall surfaces. Instead of direct modeling of volumetric and surface reactions, the proposed approach has been implemented by taking the combination of absorbing and reflecting boundary conditions at the substrate wall for DSMC particles embedded in fluid flow. This approach reduces the complexity of DSMC algorithm by simplifying surface chemical kinetics model and still obtains the local rate of adsorption of particles over wall surface.

In a broader sense, several important and widespread technologies including Microelectromechanical systems (MEMS) have relevant broad fluid flow scales ranging from meters to nanometers in frame of a single set-up [22]. The common major challenge in modeling of fluid flow in these processes is the range of fluid flow scales that requires combined use of continuum and molecular approaches [23].

The study is composed as follows. In Section 2, the DSMC code has been developed and validated using a well-known benchmark named Couette flow. The degree of gas rarefaction is represented by the Knudsen number influencing the flow velocity profile. In Section 2.1 the effect of rarefaction of CVD gas has been briefly presented. Velocity profile for a range of Knudsen numbers has been validated in Section 2.2, by comparison to prior Lattice Boltzmann results [24]. In Section 3, the FVM results of flowfield computations for the entire CVD reactor have been presented, which are used as the inlet velocity for DSMC domain. For DSMC simulations presented in Sections 3.2 and 3.3, fibers with circular cross-section in regular orientation have been computed for the range of Knudsen numbers, for varying gap between two consecutive fibers by (i) changing the volume fraction occupied by fibers and (ii) keeping the volume fraction constant . For both cases, the residence time of flow particles has been calculated which represents the amount of time the CVD feedstock gas molecules would stay inside the fibrous media. In Section 3.2, the developed in-house DSMC code is used to model particle deposition rate and residence time of particles around the substrates.

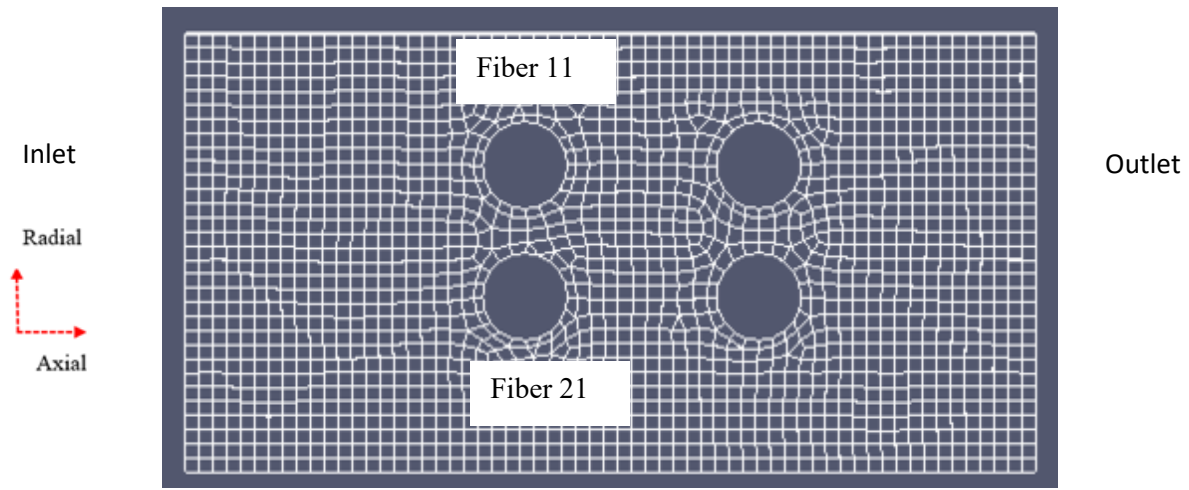

**Figure 1:** Schematic of DSMC modeling for regular fibers with DSMC cells set-up

## 2. DSMC and deposition at rigid surface

2.1 Development of DSMC and continuum models for gas flow in proximity of rigid boundary

In frame of the partial slip flow approach, 0.01 <Kn < 0.1, the continuum approximation is still valid apart from the Knudsen layer [25]. To account for the rarefaction feature at the wall, special boundary conditions are considered which are associated with the velocity slip, named Maxwell's boundary condition [26] [27].

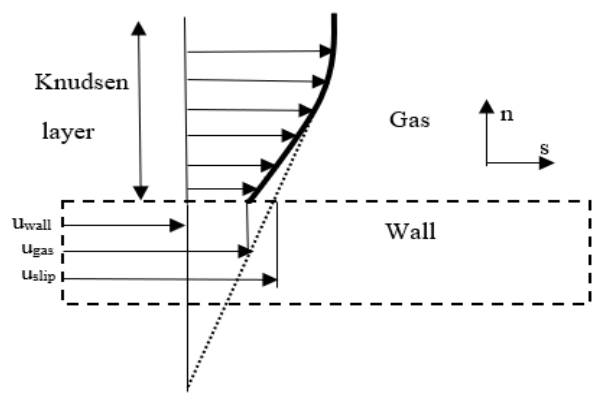

**Figure 2**: Knudsen layer and slip flow

The first-order slip velocity boundary condition for stationary wall developed by Maxwell [23,25] is used in the current study:

$$u_s = \lambda \frac{2 - \sigma_u}{\sigma_u} \frac{\partial u}{\partial n} , \qquad (2)$$

where $u_s$ is slip velocity, u is gas velocity in the Knudsen layer next to the wall, $\lambda$ is mean free path, and $\frac{\partial u}{\partial n}$ represents the slope of velocity normal to wall. The $\sigma_u$ term in Eq. (2) represents tangential accommodation coefficient. Momentum and energy transfer between gas molecules and the surface require specification of interactions between impinging gas molecules and the surface. From microscopic point of view this is quite complicated and requires detailed knowledge of scattering kernels [23] while from macroscopic point of view, the accommodation coefficient, $\sigma_u$ is accurate enough to describe the gas-wall interactions. For the limit case, $\sigma_u = 0$, gas molecules reflect from surface specularly and pre-collision and post collision velocities of gas particles are symmetric with respect to the normal direction to surface, n. For larger values of accommodation coefficient $\sigma_u$, reflections of gas molecules over the wall become more random, that represents diffusive reflection.

For flows with Kn>0.1, constitutive laws that defines stress tensor and heat flux vectors break down [28], that requires higher order corrections to the constitutive laws, resulting in Burnett or Woods equations [23], which are challenging to solve. In the current research the DSMC methodology has been implemented to compute slip velocity and to calculate the residence time of CVD gas molecules near a representative set of fibers. The DSMC approach comprises five major steps [3,16]:

1. Particle initialization.

2. Computational displacement of simulated particles after time interval $\Delta t$.

3. Indexing and cross referencing of simulated particles as they move between DSMC cells.

4. Computing of particles' new velocities after inter-particle collisions and collisions between simulated particles and wall surface.

5. Sampling the simulated DSMC particles to obtain macroscopic properties.

The first step of DSMC modeling is initialization of DSMC particles which defines the properties of fluid, DSMC cell size, number of DSMC cells and number of real gas molecules represented by DSMC particles. Simulated particles in DSMC are initialized with random velocity and random initial position. To evaluate the number of simulated particles using DSMC, let us consider a domain of volume Vc, in which molecular number density is n. The total number of real molecules in the domain is equal to nVc.

The total number of simulated particles $N = \frac{nV_c}{F_n}$, where $F_n$ is the number of real molecules represented by each simulated particle. In the current study, each simulated particle represents $10^{12}$ gas molecules. The assumption behind DSMC is that molecules have Maxwell-Boltzmann distribution everywhere in the domain and the collision between molecules are elastic in the DSMC domain. Assigned initial position and initial velocity of each DSMC particle were generated by a random number generator.

For each DSMC particle, equations of motion are solved to obtain new location of particles at the end of current, *n*-th, time step.

$$\frac{dr_i}{dt} = V_i, \quad , i = 1 \dots N \tag{3}$$

During the movement of DSMC particles, no collisions of particles are assumed.

The position of particles is updated by Eq. (4).

$$r_i^{n+1} = r_i^n + \Delta t * V_i^n \tag{4}$$

After moving DSMC particles over a time step $\Delta t$, particles are sorted and indexed. Indexing serves the purpose to figure out which particles are located in which cell. Indexing is necessary because particles might move to a next cell. The new cell locations of the molecules are indexed so that the intermolecular collisions and flow field sampling can be held accurately.

Indexing is also necessary to pick up the collision partner. In each cell, colliding particles are randomly selected. For each pair, pre-collisional velocities are replaced by post-collisional velocities for use in Eqs. (3-4) in the next time step. Proper choice of collision partner and choosing the right number of collisions are crucial to the consistency of DSMC. In the current study, the no-time-counter (NTC) method [29] has been used to model the collision. This method uses a multistep collision model. At each time moment, n, only particles, which reside in the same DSMC cell, participate in collisions. Every collision is considered as a random event with some probability of collision in every cell. The probability of collision between particles are chosen as follows

$$P_{coll}[i,j] = \left(\frac{1}{2V_c}\right)(F_N N)(N-1)(\sigma_T C_r)_{max} \Delta t \tag{5}$$

where $V_c$ is the cell volume, $(\sigma_T C_r)_{max}$ is the maximum product of collision cross section and relative speed of all possible particle pairs in the cell, $\Delta t$ is the time step size and $F_N$ is the number of real molecules or atoms that each DSMC particle represents. Particle $i$ is chosen randomly from all particles in the cell and j is chosen from the same cell to ensure near neighbor collision. The collision pair would be selected if condition below is satisfied.

$$\frac{(\sigma_T C_r)_{ij}}{(\sigma_T C_r)_{max}} > R_f. \tag{6}$$

where $R_f$ is a random number uniformly chosen at interval [0,1].

In the third stage of DSMC methodology, boundary conditions are implemented. In the current study, the model has inlet, outlet, and symmetry boundaries (see Fig. 1), which also referred to as a permeable boundary condition. For inlet, the most probable speed, U, is taken from the FVM solution for flow velocity inside the reactor. The inlet boundary is located 4 mm upstream from the DSMC domain (see Section 3 and Fig. 7). When DSMC particles exit through the upper and lower boundaries of DSMC domain (see Fig. 1), a new DSMC particle is introduced at the opposite boundary at the exit point.

To model the deposition at the substrate surface, a novel approach has been implemented by modifying the Maxwell boundary condition of diffuse scattering. A reactive coefficient or sticking coefficient [30] is introduced which represents the probability of molecules colliding with the substrate to undergo a surface reaction process, which in CVD modeling is deposition. When a simulated particle reaches substrate surface, a number $\eta$ is generated by random number generator. This number is compared with the sticking coefficient. If the value of $\eta$ is lower than the sticking coefficient, the particle would deposit in the wall. If the value is higher than the sticking coefficient, the particle would reflect to fluid domain following the cosine rule. Higher the value of sticking coefficient, the number of deposited particles would be larger. In reality, the value of sticking coefficient is a function of temperature, surface coverage, and structural details of the surface. In the current study, numerical experiments are conducted in next sections to study the effects of sticking coefficient.

In CVD process, surface reaction is the reason why particles deposit over surface, where the time available for the reaction is important. The residence time indicates how long it would take for simulated particles to move from inlet to outlet of DSMC domain. One of major objectives of the current study is to characterize the residence time of particles inside the fibrous medium for different fiber orientation set-ups and for the range of sticking coefficients at the fiber surface. If particle is reflected by the surface, the particle post reflection velocity $(v_x, v_y, v_z)$ is assigned as follows:

$$v_x = \sqrt{-2\left(\frac{k}{m}\right)T_w \log(\alpha_1)}\cos(2\pi\alpha_2), \quad (7)$$

$$v_y = \sqrt{-2\left(\frac{k}{m}\right)T_w \log(\alpha_2)},$$

$$v_z = \sqrt{-2\left(\frac{k}{m}\right)T_w \alpha \sin(2\pi\alpha_2)}$$

Here $\alpha_1$ or $\alpha_2$ are the random numbers taken from 0 to 1. In Section 3.3, the effect of the magnitude of sticking coefficient $\eta$ on the deposition rate will be explored.

2.2 Validation of DSMC method and computer code

The developed DSMC code has been used to model Couette flow for validation purpose (see Fig. 3). The Couette flow is the viscous flow between parallel infinite plates, where the upper plate is moving with uniform speed and the bottom plate is fixed. Couette flow is a shear flow driven by the upper plate movement in the flow direction. DSMC results have been validated against Lattice Boltzmann method (LBM) results developed by Refs [**Error! Bookmark not defined.**-33] for Couette flow with no-slip and partial slip boundary conditions.

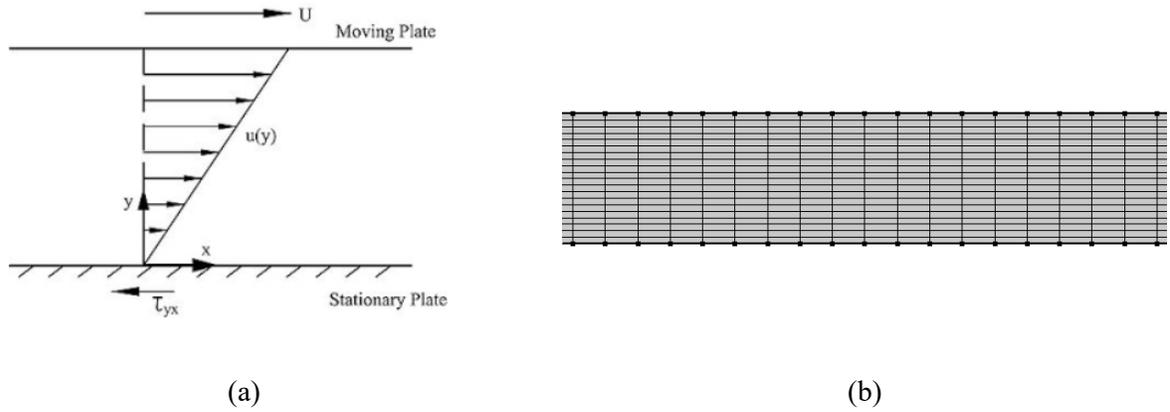

(a)          (b)

**Figure 3**: Validation of DSMC code: (a) Couette flow with no-slip boundary conditions and (b) DSMC domain with cells

For validation purpose, the distance between two plates is 1 $\mu m$ where the fluid domain is filled with Argon gas [31]. At standard temperature and pressure, the average molecular diameter is 3.66 x 10$^{-10}$ m, and the molecular mean free path is 65 nm. The plate is moving at the speed of 150 m/s (~0.5 Ma) along the positive x direction. To study the effect of Knudsen number over the velocity profile, the range of Knudsen numbers is considered and for each case the velocity profile, *u(y)*, has been computed by the DSMC code. Knudsen number in the domain can be controlled by the mean free path, pressure, and number density of gas. The velocity profile and slip velocity ae shown in Figs. 4-5 while comparison to prior LBM results is shown in Fig. 6.

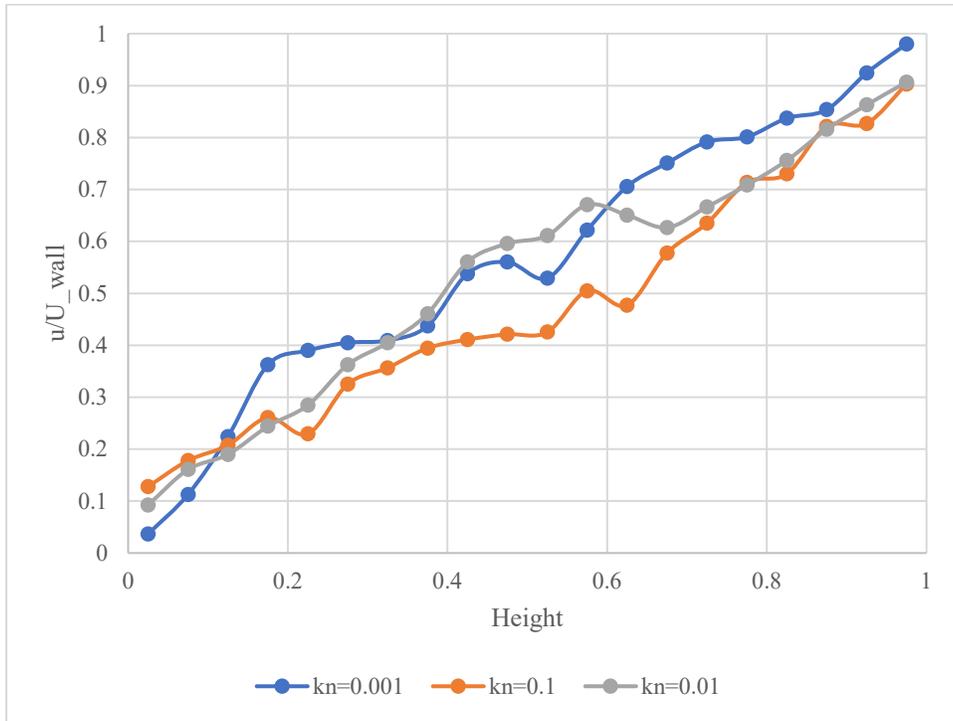

(a)

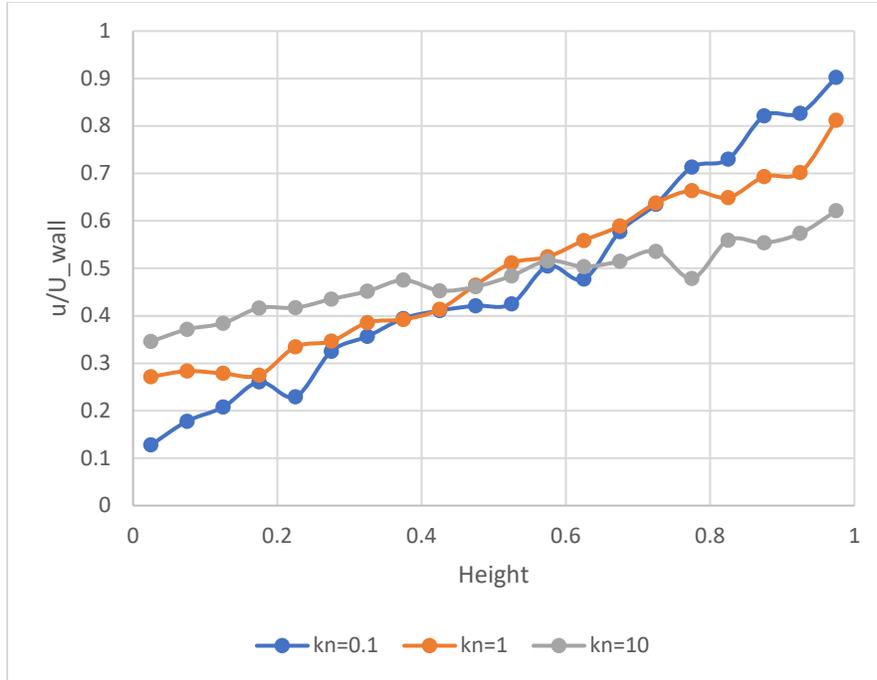

(b)

**Figure 4:** Couette flow: velocity profile for Knudsen number ranging (a) from 0.01 to 0.1 and (b) from 0.1 to 10. Velocity profile, u/U_wall, is presented as a function of coordinate y normalized by the distance b between plates.

In Fig. 4 (a) and (b), Couette flow velocity profile has been obtained for the range of Knudsen numbers corresponding to a different stage of rarefaction. The Kn > 0.01 represents partial slip flow where the velocity at the lower wall is small but not equal to zero. With increasing of the value of the Knudsen number, flow becomes more rarefied and near the bottom rigid surface the fluid velocity starts to increase. In Fig. 4 (a), for Kn=0.001 the flow velocity at the lower wall is 0.05 times the upper plate velocity while for Kn=0.1, the number increases to 0.12. In Fig. 5 (b), for Kn=1, the wall fluid velocity is 0.28 times the upper plate velocity and for Kn=10, where the flow is quite rarefied, the wall fluid velocity is 0.32 times the upper plate velocity.

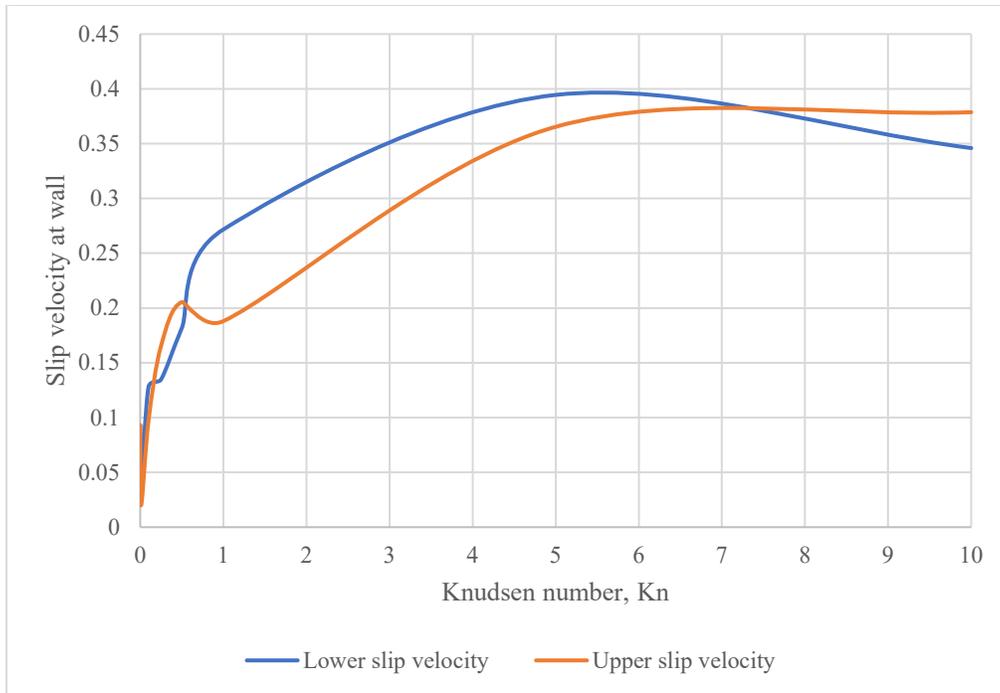

**Figure 5**: Slip velocity for Couette flow at lower and upper plates as a function of Knudsen number.

Fig 5 reveals the magnitude of slip velocity with the increase of Knudsen number. It shows that the slip velocity magnitude is quite considerable and approaches ~ 30% of upper wall speed. In this flow regime, the molecular mean free path is either comparable to or larger than the distance between plates which reduces the probability of intermolecular collisions and increase the probability of molecule-to-wall collisions. Due to higher slip velocity at both walls, the slip velocity profile flattens for Kn>0.5 compared to lower Knudsen number cases.

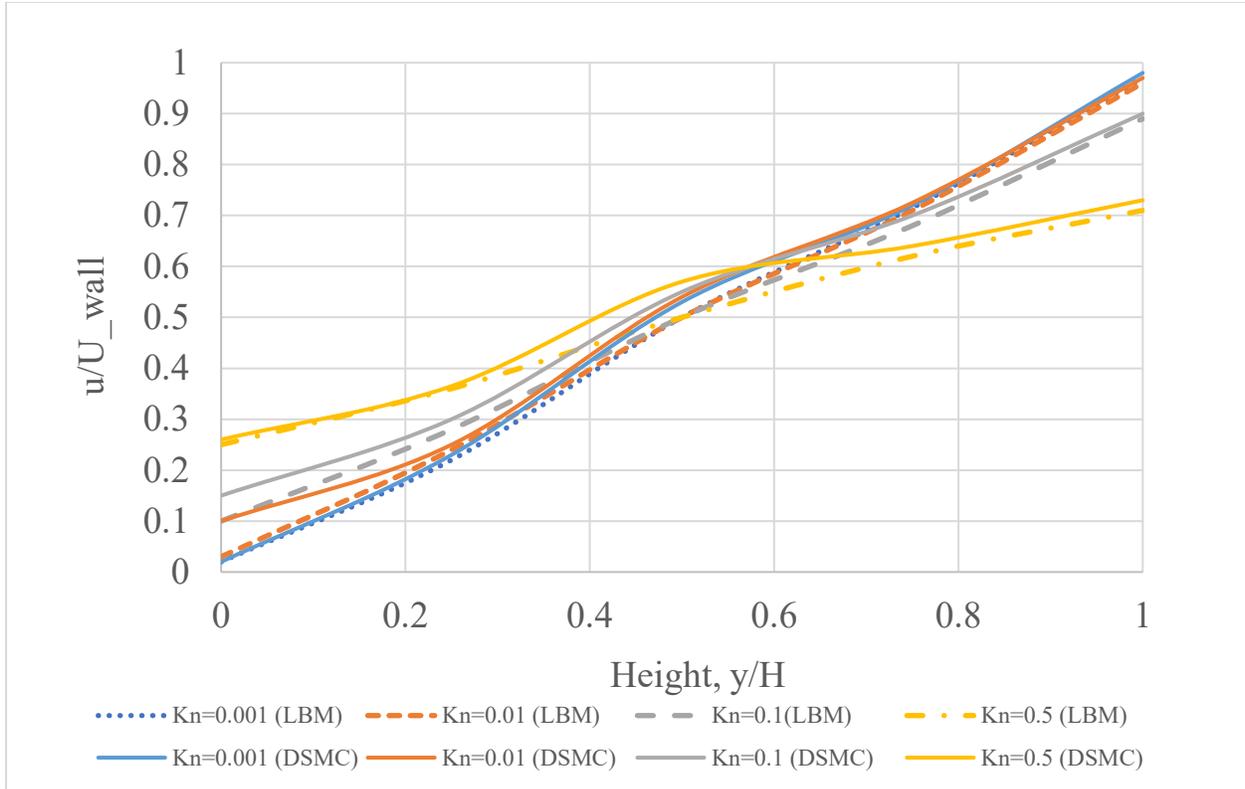

**Figure 6**: Validation of DSCM code by comparison with LBM results [24], [32], [33]

Fig. 6 shows that the DSMC results and results obtained by LBM for the same value of Knudsen number are within the acceptable error margin.

## 3. Model of deposition of carbon particles at fibers

In this section, structured sets of fibers with round-cross-section (see Fig. 1) in CVD reactor are considered.

3.1 The reactor flow field

The reactor flow field is obtained by FVM model for a CVD reactor with a bundle of fibers located at the center of reactor [34]. Pressure, temperature, and velocity are calculated by FVM at the middle of reactor, 4 mm upstream of fiber bundles and used as the inlet boundary condition for DSMC domain. By using FVM, the concentration of species in feedstock gas flow in reactor is obtained [34]. DSMC domain is encompassing the substrate (fibers) where the major feedstock gas species, which is responsible for surface CVD reaction, is $C_2H_2$.

The inlet velocity of reactor is 1.28 m/s with inlet feedstock gas temperature 25° C. The static pressure inside the reactor is 1600 Pa and the corresponding Knudsen number is 0.002 for mean free path is $\lambda = 1.87 \times 10^{-5}$ m, which represents continuum flow field. The goal of FVM simulations is to capture the feedstock gas flow velocity upstream of bundle of fibers. The velocity has been captured 4 mm upstream of the fiber bundles at the reactor centerline to be used for DSMC modeling (see Fig. 7). To justify the location of upstream boundary a parametric study has been conducted where velocity has been captured at distances of 1,2,3,4, 5,6,7,8 ,9 and 10 mm upstream of the fiber bundles. With the range of 1 mm to 3 mm upstream of the fiber bundles, velocity is lower than at locations above 3 mm due to the formation of stagnation zone over the fibers' surfaces. Beyond 4 mm upstream of fibers' bundle, the feedstock gas velocity is no longer affected by the presence of fibers. Thus, the cross-section at 4 mm upstream of the fibers' bundle is chosen to export the CVD gas velocity from FVM domain to DSMC domain.

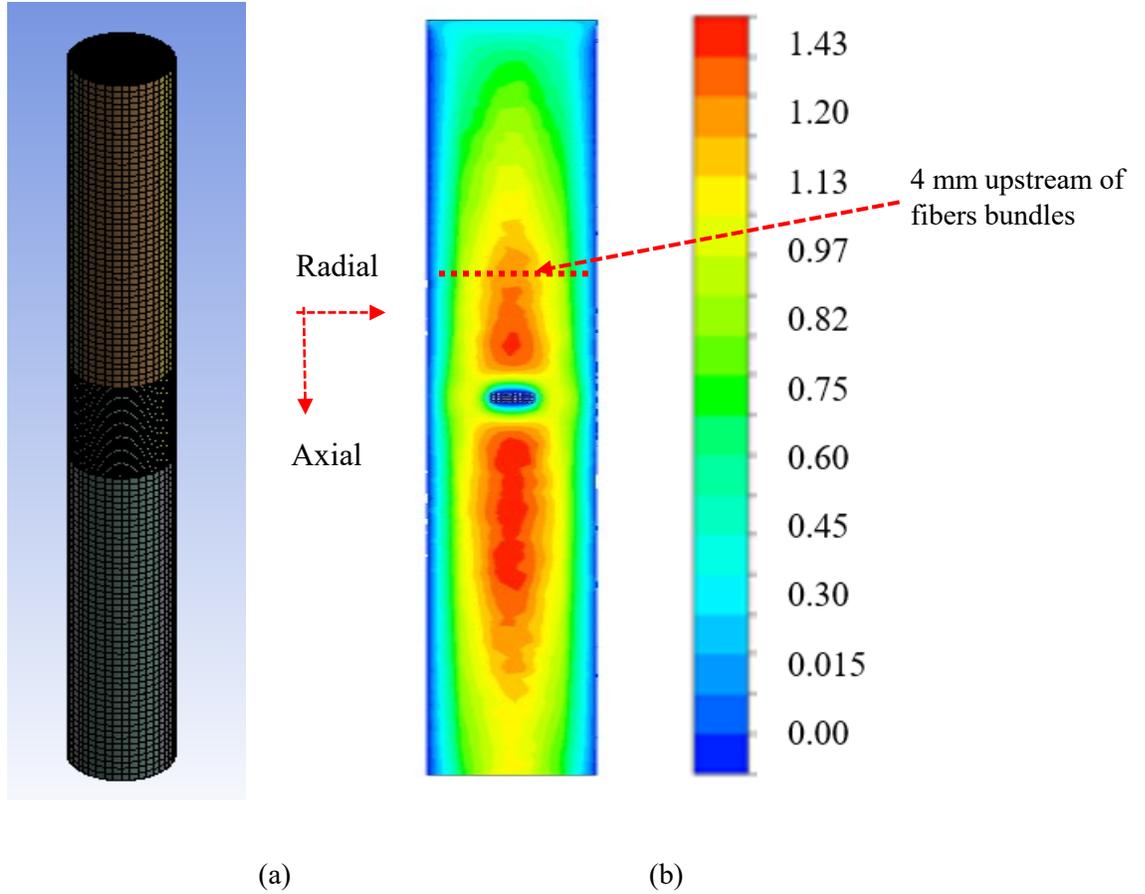

(a)                  (b)

**Figure 7:** Chemical vapor deposition reactor [43]: a) FVM mesh for CVD reactor and b) isolines of velocity magnitude along the vertical plane of the CVD reactor. The inlet boundary of DSMC domain is shown.

3.2 Set-up of fibers

In Fig. 1 the fiber set up has been shown. The diameter, $d$, of fibers is 1 μm and the molecular mean free path of the gas is $2.5 \times 10^{-07}$ m which corresponds to Kn=0.25. Recall that the range Kn > 0.1 corresponds to the rarefied zone. The DSMC domain length is 10 μm (10d) while the domain height is 6 μm (6d). For the naming convention of fibers in Fig. 1, the first number denotes the row of fibers and the second number denotes the column of fibers. So, "Fiber21" means fiber belonging to the $2^{nd}$ row and $1^{st}$ column in the bundle.

The size of fibers' domain is much smaller compared to the length of reactor. The end effect of the fibers can be neglected because the fiber length is much larger than fibers' diameter. Symmetry can be considered along the Z axis to account for multiple fibers in the bundle. The side surfaces are taken as symmetry boundaries to account for the fact that the bundle structure is self-repeating.

In DSMC modeling (see Section 2.1), the first step is to populate the domain with simulated particles where each simulated particle represents billions of real molecules. The inlet velocity of DSMC domain is imported from the FVM computations and is equal to 0.97 m/s for current computations (see Fig. 7). Particles are initialized with the inlet velocity and allowed to move from domain inlet to domain exit. At the fibers' surface, combined diffuse and reflective boundary condition has been considered, where the probability of reflection is *p* and probability of sticking to the fiber surface is *1-p* (see Section 2).

.

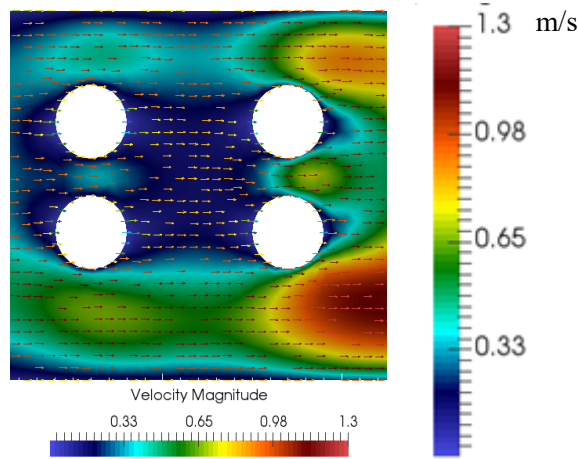

**Figure 8**: Velocity field around fibers in bundle computed by DSMC

Figs. 8 shows that in the interior of the array of fibers in a bundle, fluid velocity magnitude is smaller, ~0.33 m/s, compared to the inlet velocity of ~0.97 m/s. Between two fibers, velocity increases due to the partial blockage of flow area as follows from the conservation of mass.

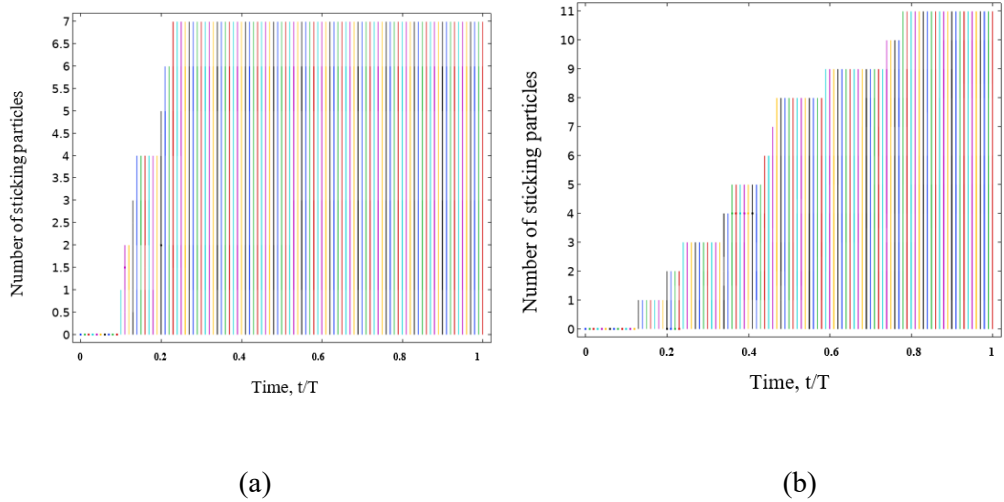

(a)                          (b)

**Figure 9:** Particle deposition intensity over time at fiber surface: (a) Fiber 11 (Row 1, Column 1) and (b) Fiber 12 (Row 1, Column 2) for 50% sticking coefficient.

The normalized time, t/T, is used, where T=10 μs

Fig 9 shows the intensity with which simulated particles deposit at the fibers' surface. Particles, which flow from the inlet of DSMC domain, start to deposit at an individual fiber surface after particles reach it. Comparison of Fig. 9(a) to Fig. 9(b) shows that the front row fiber reaches the steady state in terms of deposition rate more rapidly comparing to the second-row fiber. While after 4 μs the number of deposited particles does not change with respect of time (Fig. 9a), for the second row fiber it takes ~ 8 μs (Fig. 9b). The steady state number of deposited particles is different by almost 50% between the first and the second row fibers (Table 1).

Table 1 Number of deposited particles over fiber surfaces

| Fiber | Number of Simulated particles | Mass of deposited layer, nanogram |
|---|---|---|
| Fiber 11 | ~7 | 290 |
| Fiber 12 | ~11 | 457 |
| Fiber 21 | ~7 | 290 |
| Fiber 22 | ~11 | 457 |

Table 1 shows how carbon deposits over fiber surfaces for regular orientation of fibers (Fig.1 ) with Kn=0.3. Molar mass of $C_2H_2$ is 26.04 g/mol, after subtracting 2 atoms of $H_2$, as hydrogen does not participate in surface deposition , the molecular mass of remaining $C_2H_2$ is 4.16 x $10^{-23}$ kg. As each DSMC particle consists of $10^{12}$ molecules, the mass of each DSMC particle would be 4.16 x $10^{-11}$ kg.

3.3 Effect of sticking coefficient on deposition rate

In the previous section, simulations were conducted with the sticking coefficient equal to 50%. To understand the role of the value of sticking coefficient, a parametric study has been conducted in the current section for a range of sticking coefficient, $\eta$, namely, 20%, 40%, 50%, 60% and 80%. With a higher value of sticking coefficient, particles would have higher probability to stick to the fiber surface while a lower probability of sticking repel more particles away from the fiber surface. The simulations conducted in the current section quantify the trend.

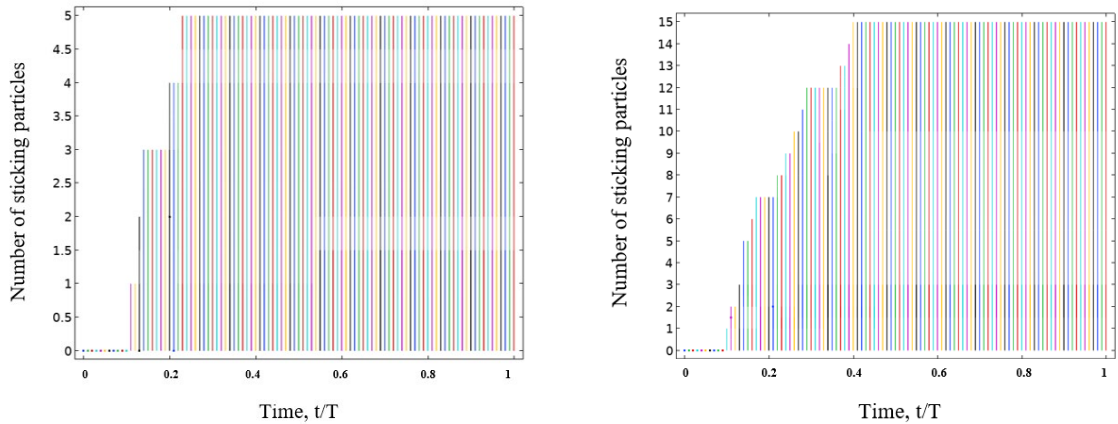

**Figure 10**: Carbon particle deposition on fiber 11 (First row, first column fiber) for (a) 20% sticking coefficient and (b) 80% sticking coefficient

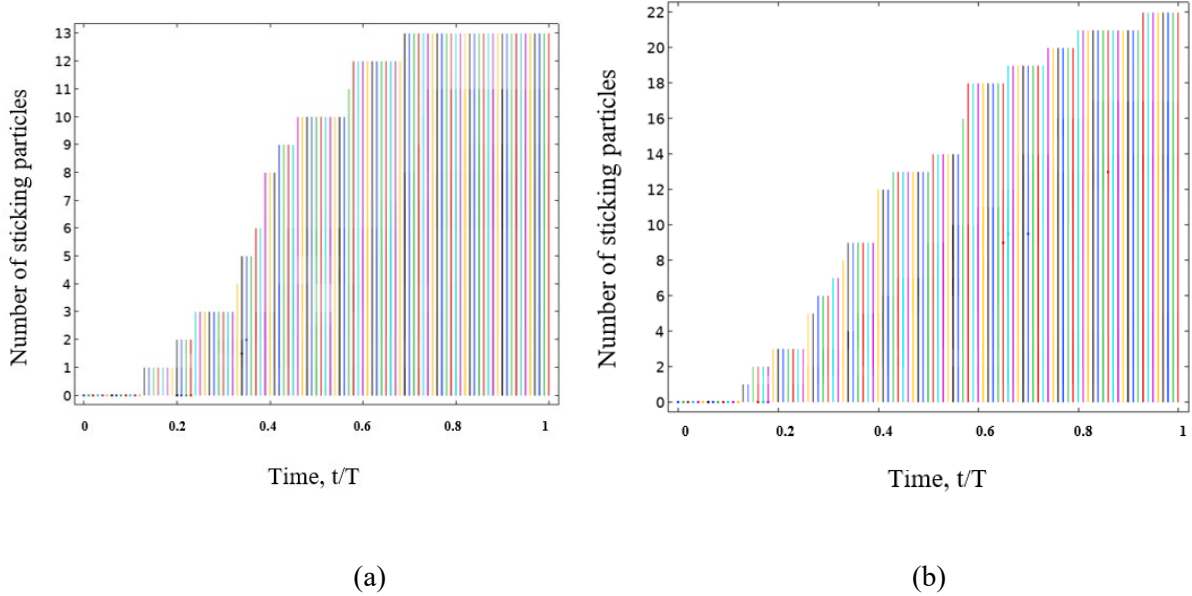

(a)                  (b)

**Figure 11**: Carbon particle deposition on fiber 12 (First row, second column fiber) for (a) 20% sticking coefficient and (b) 80% sticking coefficient

Fig. 10 and 11 show how the particle deposition rate varies with respect to the value of sticking coefficient and with respect to position of fiber in bundle. Front row fibers, which face the flow stream first, have lower deposition rate compared to the deposition at surface of second-row fibers. Generally speaking, higher sticking coefficient provides higher deposition rate while lower sticking coefficient provides lower deposition rate. This dependence is monotonic and non-linear as detailed in Fig. 12.

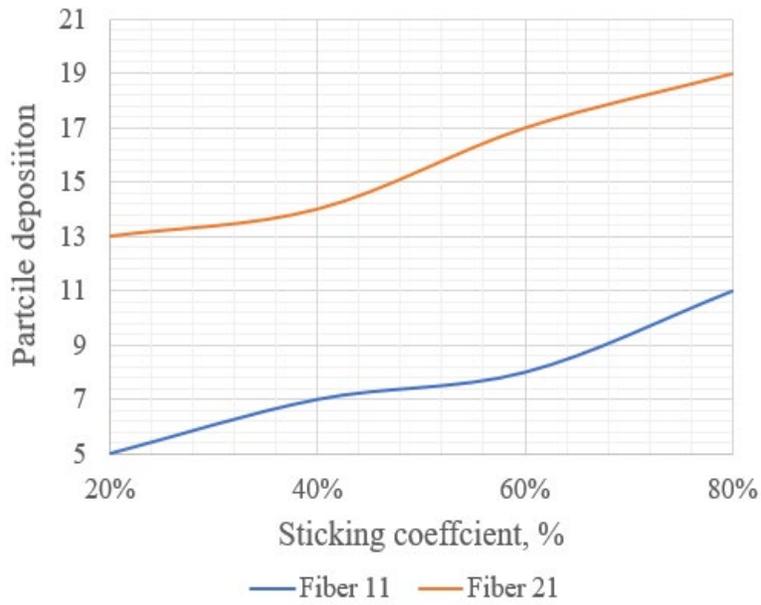

**Figure 12:** Particle deposition as a function of sticking coefficient at an individual fiber surface

3.4 Effect of distance between the fibers

The intermittent distance between the fibers plays a vital role in deposition magnitude and profile. It will be shown in this section that for the same volume fraction of fibers in bundle, the closer the distance between fibers the larger the particles' deposition.

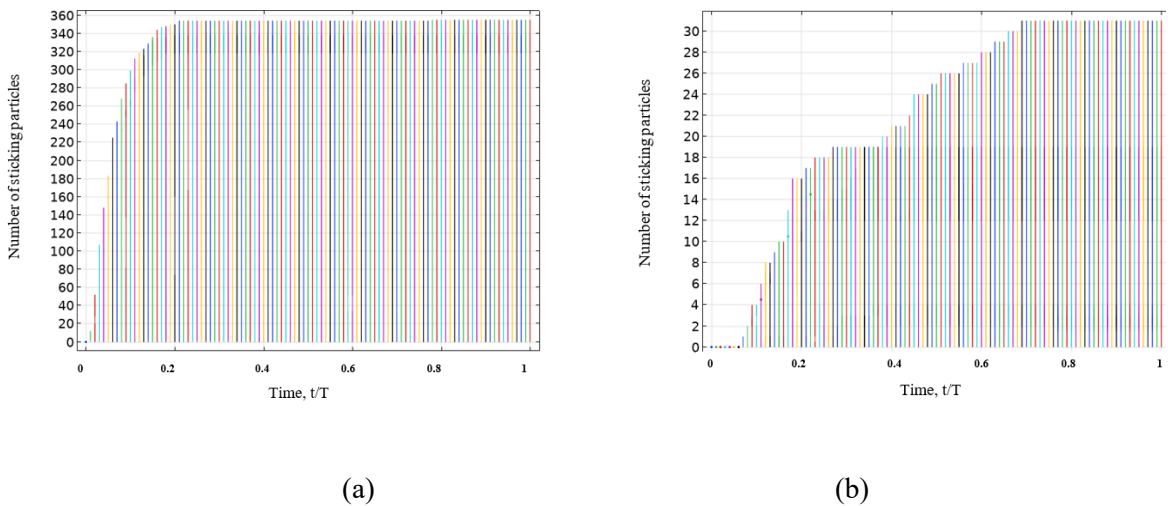

(a)                                                                 (b)

**Figure 13**: Particle deposition rate for 0.25 mm diameter fiber at 0.5 mm between fibers:

(a) Fiber 11 (Row 1 and Column 1) and (b) Fiber 12 (Row 1 and Column 2)

The volume fraction of fibers is 20% and sticking coefficient is 20%

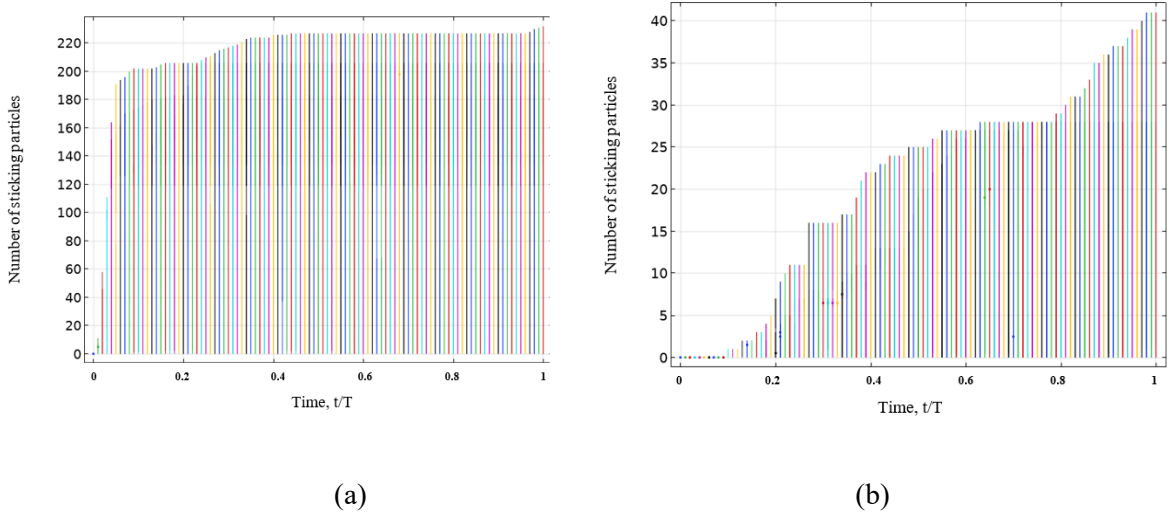

(a)   (b)

**Figure 14:** Particle deposition rate for 0.5 mm diameter fiber at 1 mm between fibers

(a) Fiber11 and (b) Fiber12. The volume fraction and sticking coefficient are the same as in Figure 13.

Figs.13 and 14 show that for the 20% volume fraction of fibers in bundle and 20% sticking coefficient at fiber wall surfaces, the number of deposited particles varies considerably with respect to time. For 0.5 mm distance between centers of 0.25mm in diameter fibers, for time moment t/T=0.25 ~360 simulated particles have deposited at the fibers' surface while for the same 20% volume fraction, with 1 mm gap and 0.5 mm diameter, at 40% time, only 30 particles have deposited. It shows that smaller gap between the fibers and smaller diameter of fibers causes ~12 times larger deposition for front row fibers. For the second row fibers, the deposition gain is ~5 times.

(a)

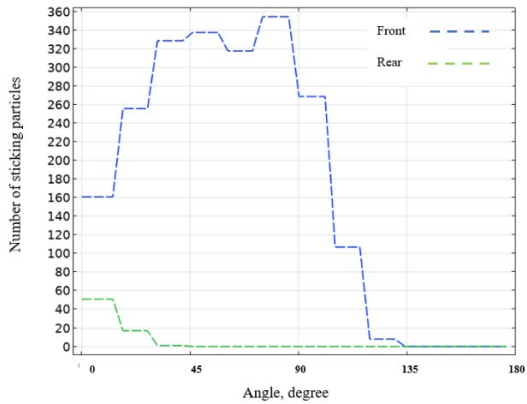 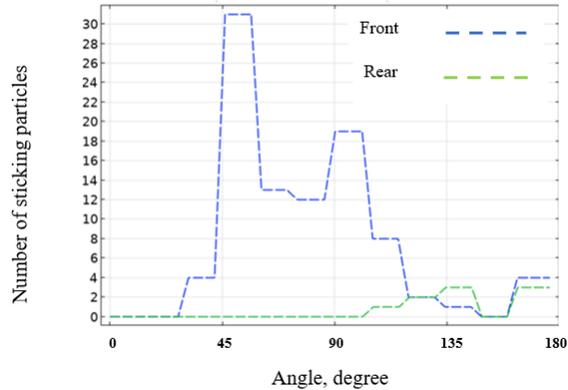

(b)                  (c)

**Figure 15**: Particle deposition profile at Fiber 11: (a) fibers of 0.25 mm diameter and 0.5 mm distance and (b) fibers of 0.5 mm diameter and 1 mm distance.

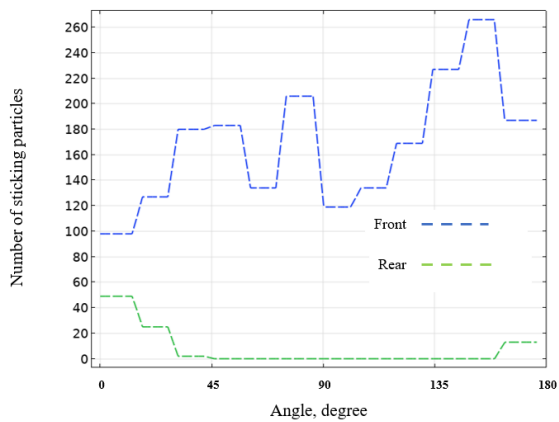 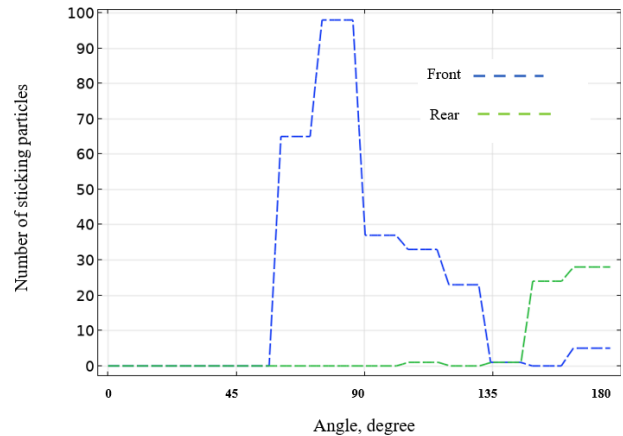

**Figure 16:** Particle deposition profile at Fiber 12: diameters of fibers and distance between fibers are the same as in Fig. 15 (a, b).

Figs. 15 and 16 a show that the deposition profile at the front (facing the flow) and rear (facing the separation zone) sides of fibers. The deposition profile in Figs. 15 and 16 is obtained for the same volume fraction of fibers in bundle. For both cases, the front side of the fiber shows higher deposition than that for the rear side. For smaller gap and diameter, the majority of deposition occurs between 0° and 135° while for the rear fiber, it occurs between 0° and 45° and between 135° and 180°. With larger fibers and larger distance between them (Figs. 15(b) and 16 (b)), the deposition is more pronounced within the 45°
to 135° degrees range.

3.5 Effect of rarefaction

Rarefaction in gas affects the deposition profile at fibers' surface. Rarefaction happens either due to lower pressure or due to smaller diameter of fibers. With a higher Knudsen number (Eq. (1)) caused by smaller fiber diameter with the same volume fraction of fibers in bundle, gas particles have to travel longer to collide due to a higher molecular mean free path. Also, the slip velocity increases with the Knudsen number. With a higher slip velocity, flow resistance is lower due to a smaller velocity gradient normal to rigid surface and, consequently, reduced skin friction. To quantify the residence time of particles in the vicinity of fibers, an area of influence (Fig. 17a) has been considered around the bundle of fibers.

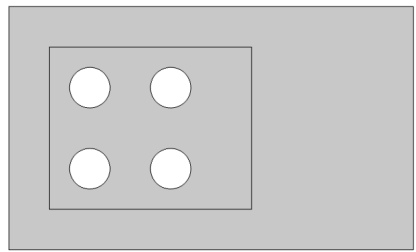

(a)

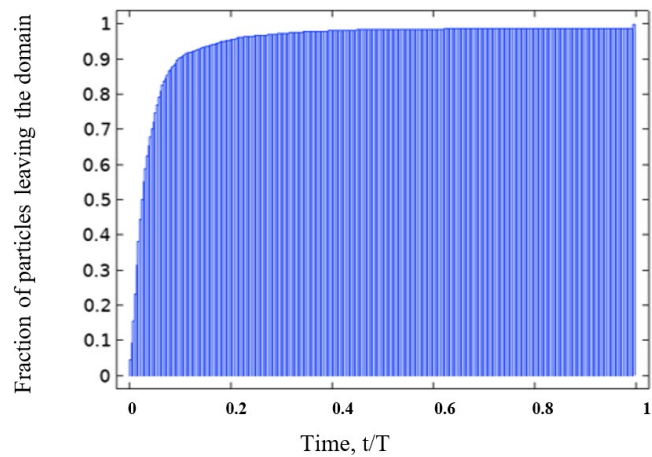

(b)

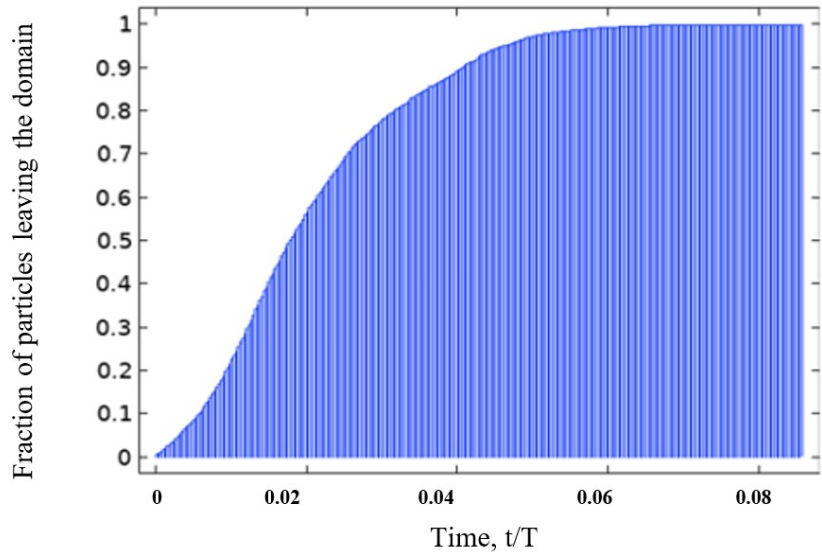

(c)

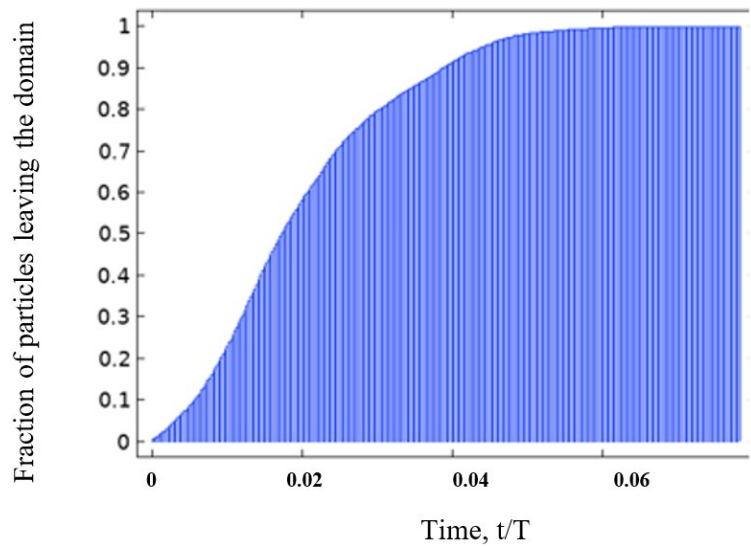

(d)

**Figure 17:** Residence time for particles in flow across the set of fibers for sticking coefficient equal to 50% for the range of Knudsen numbers: (a) area of influence around the bundle of fibers; (b) no-slip condition; (c) Kn=0.3 and (d) Kn=0.5. Hear T=10 μs.

For no-slip condition (Fig. 17 (b)), at the surface of fibers all particles which were not sticking at the fibers' surface leave the domain within 2 microseconds (0.2 x 10 µs). For Kn=0.3 (Fig. 17 (c)), where the gas flow is partially slip at fibers' surface, the residence time is smaller due to lower friction resistance at the fibers surface. Particles, which do not stick in the fiber surface, leave the domain within 0.7 x 10 µs=7 µs . For Kn=0.5(Fig. 17 (d)), the residence time decreases to 0.6 x 10 µs = 6 µs. With a lower residence time, available time for particles to participate in surface deposition decreases and the amount of deposition would be lower compared to no-slip conditions.

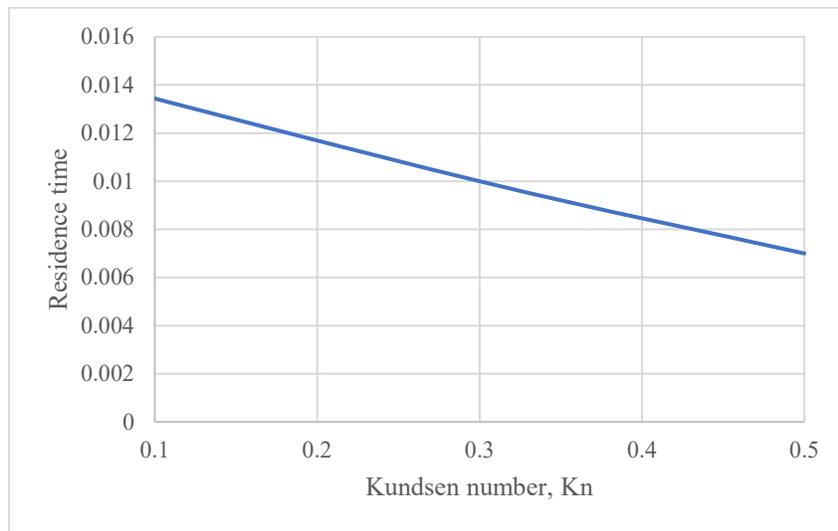

**Figure 18**: Residence time vs Knudsen number

Fig. 18 shows that with the increase of Knudsen number, the residence time decreases nearly linearly. In terms of physical significance, with smaller fiber size the residence time would be smaller due to slip velocity at wall which would offer less time for surface reactions. With less time available for surface reaction, CVD process, which is relatively slow process itself, would produce less solid deposition in substrate surface.

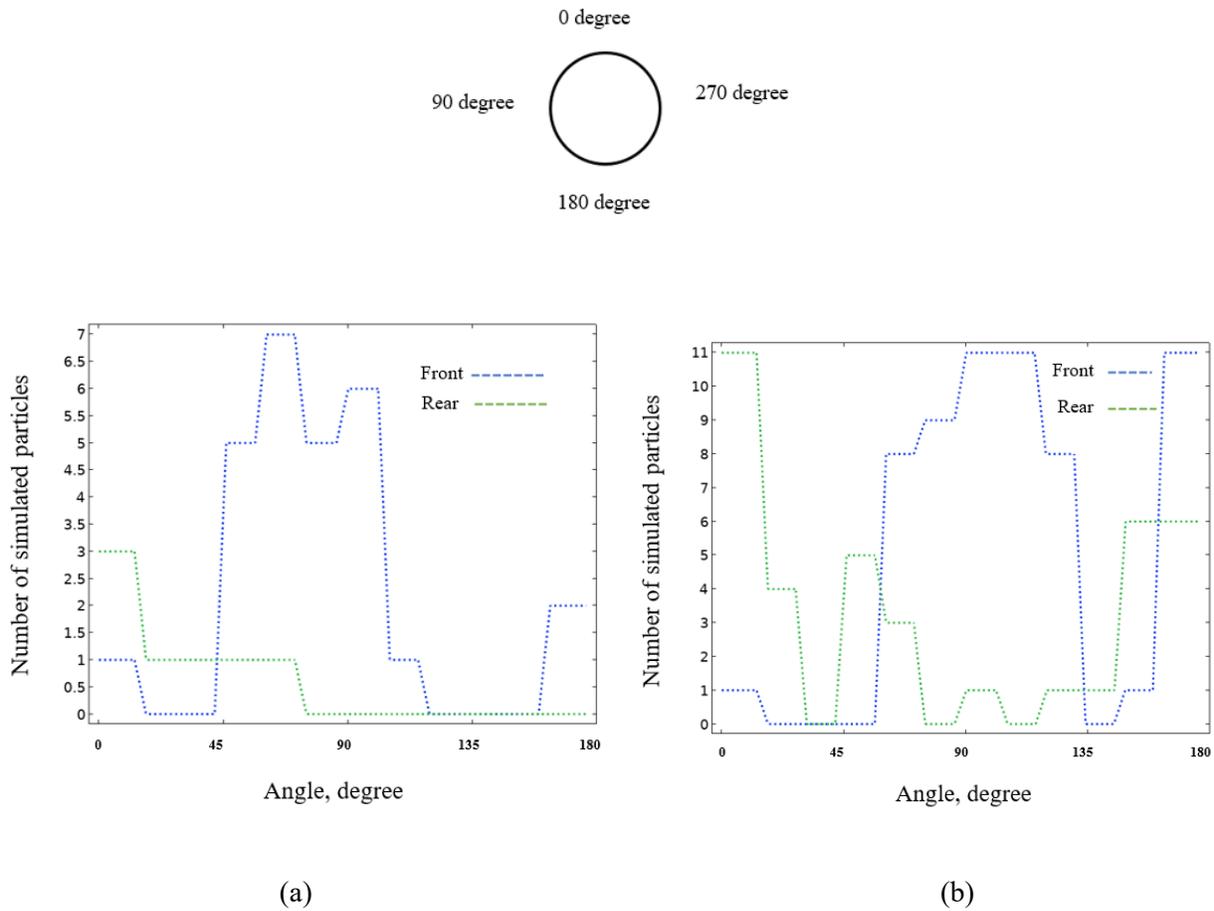

(a)                                                                (b)

**Figure 19:** Particle deposition profile at (a) Fiber 11 and (b) Fiber 12 for Kn=0.3

Fig.19 (a) shows the deposition profile along the fiber circumference modeled by DSMC method. In the front side of fiber with respect to flow (for angles between 45° and 135°), deposition is larger compared to the rear side for Fiber 11. For Fiber 12, both front and rear semi-circumferences have larger deposition of particles compared to Fiber 11. Due to the low-speed wake downstream of Fiber 11 (see Fig 8), CVD gas has larger residence time near the second fiber (Fiber 12) that allows larger and more uniform deposition compared to the first row fibers.

## Conclusions

The goal of this study is to adopt Direct Simulation Monte Carlo (DSMC) to evaluate the CVD rate for fibers of nano-scale diameter by using simulated $C_2H_2$ particles. For micro- and nano- scale fibers with non-zero Knudsen number, conventional finite volume method (FVM) is not applicable while molecular methods like DSMC are very computationally intensive to model flowfield for entire industrial reactor. This study couples FVM with DSMC, where bulk flow in reactor was computed by FVM and flow near the fibers is modeled by DSMC.

With reduction of fiber diameter, the corresponding Knudsen number increases and velocity at the fibers' surface starts to deviate from no-slip to slip velocity mode. Due to slip velocity at fibers' surface, the residence time of particles representing reactive molecules within the fibrous medium decreases that eventually decreases the amount of surface deposition at the fibers' surface. The deposition magnitude appears to be a strong function of the sticking coefficient. For the 80% sticking coefficient, the deposition rate is nearly doubled compared to that for the 20% sticking coefficient. Fibers at front row show higher deposition rate compared to fibers at the second row due to longer contact with feedstock gas compared to the second row fibers.

In terms of angular position at fiber surface, for the direct exposure to the flow at the front stagnation point, the deposition is the highest for front row fibers. For second row fibers, maximum deposition occurs near equatorial of the circumference of fiber, where wake starts to form.

The proposed methodology will be extended to unstructured set-ups of fibers with multiple diameters in the future research.

[17] Zhao, S., and Povitsky, A., 2013, "Coupled Continuum and Molecular Model of Flow through Fibrous Filter," *Phys. Fluids*, **25**(11).

[18] Dreyer JAH, Riefler N., Pesch GR, Karamehmedovic M., Fritsching U., Teoh WY, and Madler L Simulation of Gas Diffusion in Highly Porous Nanostructures by Direct Simulation Monte Carlo. *Chemical Engineering Science*, 2014, Vol. 105. pp. 69–76.

[19] Pesch, G. R., Riefler, N., Fritsching, U., Ciacchi, L. C., and Madler, L., "Gas-Solid Catalytic Reactions with an Extended DSMC Model," *AIChE Journal* 61(7), 2015.

[20] Bird, G. A., The DSMC Method, Version 1.2, 2013.

[21] Bird, G. A., 2011, "The Q-K Model for Gas-Phase Chemical Reaction Rates," *Phys. Fluids,* **23**(10), pp. 1–13.

[22] "What Is MEMS Technology?" [Online]. A MEMS Clearinghouse® and information portal for the MEMS and Nanotechnology community, https://www.memsnet.org/mems/what_is.html. [Accessed: 23-Dec-2021].

[23] Karniadakis, G., Beşkök, A., and Aluru, N. R., 2005, "Microflows and Nanoflows : Fundamentals and Simulation," Springer.

[24] Shirani, E., and Jafari, S., 2007, "Application of LBM in Simulation of Flow in Simple Micro-Geometries and Micro Porous Media," *African Physical Review* (2007) 1:0003.

[25] P. A. Chambre and S.A. Schaaf, 1961, "Flow of Rarefied Gases", ISBN: 9780691654904, Princeton Univ. Press. Published: Mar 21, 2017

[26] E. H. Kennard , "The Kinetic Theory of Gases. With an Introduction to Statistical Mechanics", 1st Ed. London: McGraw-Hill Publishing Co., Ltd., 1938; also see Review, *Journal of Chemical Technology and Biotechnology*, **57**(39), 1938, p. 901.

[27] Colin, S., 2012, "Gas Microflows in the Slip Flow Regime: A Critical Review on Convective Heat Transfer," *J. Heat Transfer*, **134**(2), p. 020908.

[28] S. Chapman, and T. G. Cowling, "The Mathematical Theory of Non-Uniform Gases", 1953.

[29] E. S. Oran, C. K. Oh, and B. Z. Cybyk, "Direct simulation Monte Carlo: Recent advances and applications," *Annu. Rev. Fluid Mech*. **30**, 403 (1998).

[30] Y. Akiyama, S. Matsumura, and N. Imaishi, Shape of film grown on microsize trenches and holes by chemical vapor deposition: 3-dimensional Monte Carlo simulation, *JPN. J. Appl. Phys*. 34 (11) (1995) 6171-6177.

[31] A. Garcia, "Numerical Methods for Physics (Python) ", Second, Revised (Python) Edition, 2017.

[32] Lockerby, D. A., Reese, J. M., and Gallis, M. A., 2005, A Wall-Function Approach to Incorporating Knudsen-Layer Effects in Gas Micro Flow Simulations. *AIP Conference Proceedings 762, 731*

[33] Niu, X. D., Shu, C., and Chew, Y. T., 2004, "A Lattice Boltzmann BGK Model for Simulation of Micro Flows," *Europhys. Lett*., **67**(4), pp. 600–606.

[34] Barua, H., and Povitsky, A., 2020, "Numerical Model of Carbon Chemical Vapor Deposition at Internal Surfaces," *Vacuum*, **175**, p. 109234.